\documentclass[14pt,osajnl2,preprint,showpacs,superscriptaddress]{revtex4}
\usepackage[draft]{hyperref}
\usepackage{graphicx}
\usepackage{amsfonts,amssymb,amsmath,bm,textcomp}
\usepackage[dvips]{color}
\usepackage{times}

\begin{document}

\title{Localized surface plasmon modes in a system of
%\textcolor{red}
{two} interacting metallic cylinders}

\author{Viktoriia E. Babicheva}
\address{Moscow Institute of Physics and Technology,  Institutskij  pereulok 9, 141700 Dolgoprudnyj, Moscow Region, Russia}
\address{DTU Fotonik, Technical University of Denmark, Oersteds Plads 343, 2800 Kgs. Lyngby, Denmark}
\author{Sergey S. Vergeles}
\address{Moscow Institute of Physics and Technology,  Institutskij  pereulok 9, 141700 Dolgoprudnyj, Moscow Region, Russia}
\address{Landau Institute for Theoretical Physics RAS, Kosygina 2, 119334 Moscow, Russia}
\author{Petr E. Vorobev}
\address{Moscow Institute of Physics and Technology,  Institutskij  pereulok 9, 141700 Dolgoprudnyj, Moscow Region, Russia}
\address{Landau Institute for Theoretical Physics RAS, Kosygina 2, 119334 Moscow, Russia}
\author{Sven Burger}
\address{Zuse Institute Berlin, Takustra\ss e 7, 14195 Berlin, Germany}

\begin{abstract}
We study an optical response of a system of two parallel close metallic cylinders
having nanoscale dimensions.
Surface plasmon excitation in the gap between the cylinders are specifically analyzed.
%\textcolor{red}
{In particular, resonance frequencies and field enhancement were investigated as
functions of geometrical characteristics of the system and Ohmic losses in the metal.
The results of numerical simulations were systematically compared with the analytical theory,
obtained in the quasi-static limit.
The analytical method was generalized in order to take into account the retardation effects.}
We also present the physical qualitative picture of the plasmon modes, which is validated
by numerical simulations and analytical theory.
\end{abstract}

%\ocis{160.3918, 240.6680, 260.3910.}

\maketitle

\section{Introduction}

Localized surface plasmon excitations in metal-dielectric systems of subwavelength size is
a topic under intensive study during the last decade.
Fabrication techniques of wire-grid polarizers \cite{Suzuki2010,Ekinci2006},
nano-antennas \cite{Berthelot2009,Bakker2008,Lakowicz2007,Bloemendal2006,Jain2007},
and arrays of metallic particles \cite{Zhou2010,Sanders2006,Maier2003} are rapidly improving.
Optical properties of aggregates of metallic grains are
very different from those of the separate grains.
Field refraction by such systems possesses typical features,
in particular a strong field enhancement in the gaps between closely located metallic particles leading to
an increase of scattering and absorption in comparison with those for single grains.
The resonance frequency of surface plasmons in the systems
depends both on particle sizes and inter-particle distances.
It is red-shifted for modes with electric field polarization
directed across the gaps between the grains
and blue-shifted in the opposite case,
see e.g. experimental  works   \cite{Bloemendal2006,Jain2007}
and numerical investigations \cite{Clarkson2011,Amendola,Romero2006}.
The electric field enhancement inside the gaps under the resonance conditions can reach sufficient values
and one can use the effect to achieve Raman detection of single molecules
placed into the gap \cite{Cheng2011,Haran2010}.

One encounters difficulties trying to describe plasmon modes analytically since an exact
solution is only possible for systems with very simple geometry. The geometry should allow to use an appropriate coordinate
system for which separation of variables in the Helmholtz equation is possible \cite{Meshbach}.
Among the systems are spherical metal particles \cite{Boardman1977},
surface plasmon propagation in plane metal films \cite{Fedyanin2010a},
dielectric gaps in metallic cladding \cite{Kaminov-Mammel_1974_ApplOpt}
and along nanowires \cite{Pfeiffer1974}.
Although the solutions for these systems are quite simple, they yield
basic understanding of the fundamentals of the surface plasmon physics at nanoscales.

%\textcolor{red}
{The problem of two particles (in particular cylinders) in the external field can be
approximately solved using dipole-dipole approximation if the inter-particle distance is much larger than their sizes.
When the inter-particle distance gets smaller this approximation becomes inappropriate even
for qualitative description of the system, and one should use multipole-multipole
expansion technique \cite{Stockman2004}.
There are some works that try to employ the multipole-multipole
expansion technique in order to approach a system of two close metallic particles of more complex form \cite{Zhukovsky2011},
system of several particles \cite{Hentschel2011} or metamaterials \cite{Petschulat2008,Chigrin2011}.
which is associated with the calculation of formally infinite series.
The method has advantages for numerical simulation,
whereas it does not allow to establish qualitative properties of the plasmon modes in the systems,
in particular their scaling behavior on the geometry of the system.}

The problem can be simplified at scales less than the wavelength.
In this case the Helmholtz equation is reduced to Laplace equation
since retardation effects are negligible.
This fact allows one to analytically solve the problem about
surface plasmon mode structure for more complex systems such as
two close spherical grains \cite{Lebedev,Klimov2007} and two close
cylinders with circular cross section \cite{Vorobev}.
To analyze solutions it is reasonable
first to consider a qualitative picture \cite{Lebedev,Vorobev} which
describes the plasmon modes in the systems.
The qualitative picture can be constructed applying the solution for surface plasmon propagation in a thin dielectric gap
\cite{Kaminov-Mammel_1974_ApplOpt} to the gap between two granules.

In the present work, we investigate scattering of light by a system of two close parallel metallic
cylinders. We sequentially compare the results
of numerical simulations of the electromagnetic near field
distribution with predictions of the
theory \cite{Vorobev} developed for the quasi-static (long wavelength)
limit. Following \cite{Vorobev} we present the qualitative picture
of the surface plasmon resonance in the system in more details, and
show its agreement with both analytical (in quasi-static limit)
and numerical solution of Maxwell's equations.

%\textcolor{red}
{One of the main flaws of the method employed in \cite{Vorobev} is
that it does not account for retardation effects, which means that its applicability
diminishes with the increase of the system size. Moreover, it does not account for the
radiation losses which could be significant even for the small system provided that the Ohmic
losses are small enough. In the present paper, and it is its main point, we present the numerical
results which on the one hand account for the retardation effects, thus are applicable for the systems
of any size and material constants, and on the other hand extend the analytical methods of \cite{Vorobev}
to account for the radiation losses explicitly. We verify the accuracy of our numerical calculations by systematically
comparing their results to analytical ones in the appropriate limit of small system size. We also present the results of
numerical simulations for the silver cylinders in the experimentally interesting region of frequencies and
for realistic sizes (up to 100 nm) which is slightly beyond the scope of the analytical quasi-stationary
method which describes the picture only qualitatively.} Silver is chosen since it is widely used in in experimental works
in nanooptics (see e.g. \cite{Michaels2000,Maier2003}) due to its low Ohmic losses.

\section{Problem formulation}

\begin{figure}[t9]
\includegraphics[width=0.70\textwidth]{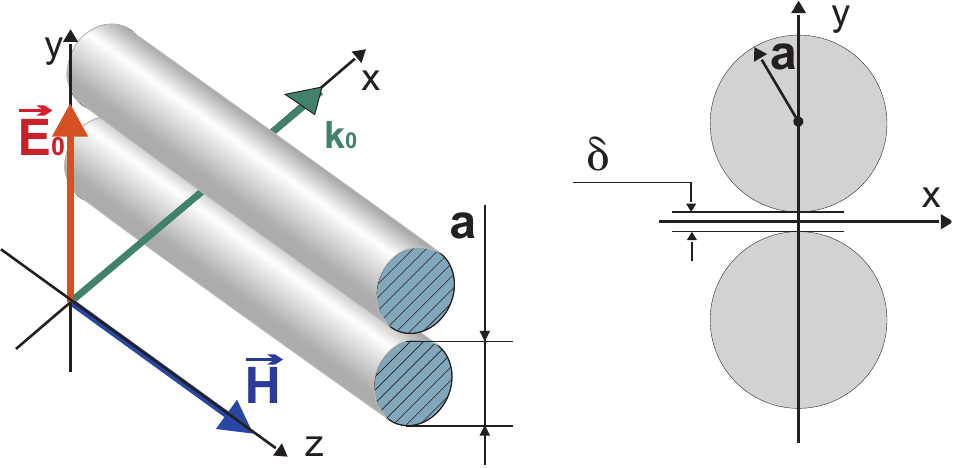}
\caption{Two close cylinders, coordinate system and field polarization.}
\label{chain2}
\end{figure}

We examine scattering of an electromagnetic plane wave by two close metallic
cylinders under the conditions of surface plasmonic resonance.
The size of the cylinders in cross-plane is assumed to be of the order or less
than the wavelength. FIG. \ref{chain2} illustrates the system of two
cylinders with circular cross sections together with the Cartesian
coordinate system that we operate.
The radii of both cylinders are equal to $a$ and the width of the gap between the cylinders is $\delta$.
We consider a particular case of
TM-wave when the electric field of the incident wave is polarized along the line
connecting the cylinders axes.
Such choice is made because surface plasmon modes with red-shifted resonance frequencies
and high local enhancement of the electric field in the
gap between the cylinders \cite{Vorobev} is achieved for this polarization only.
The red shift corresponds
to high negative value of the dielectric constants contrast (permittivity ratio)
$\varepsilon = \varepsilon_\mathrm{m}/\varepsilon_{d}$, where
$\varepsilon_\mathrm{m}$  and $\varepsilon_\mathrm{d}$ are the
permittivities of the metal and the surrounding
dielectric respectively.

\begin{figure}[t9]
\includegraphics[width=0.8\textwidth]{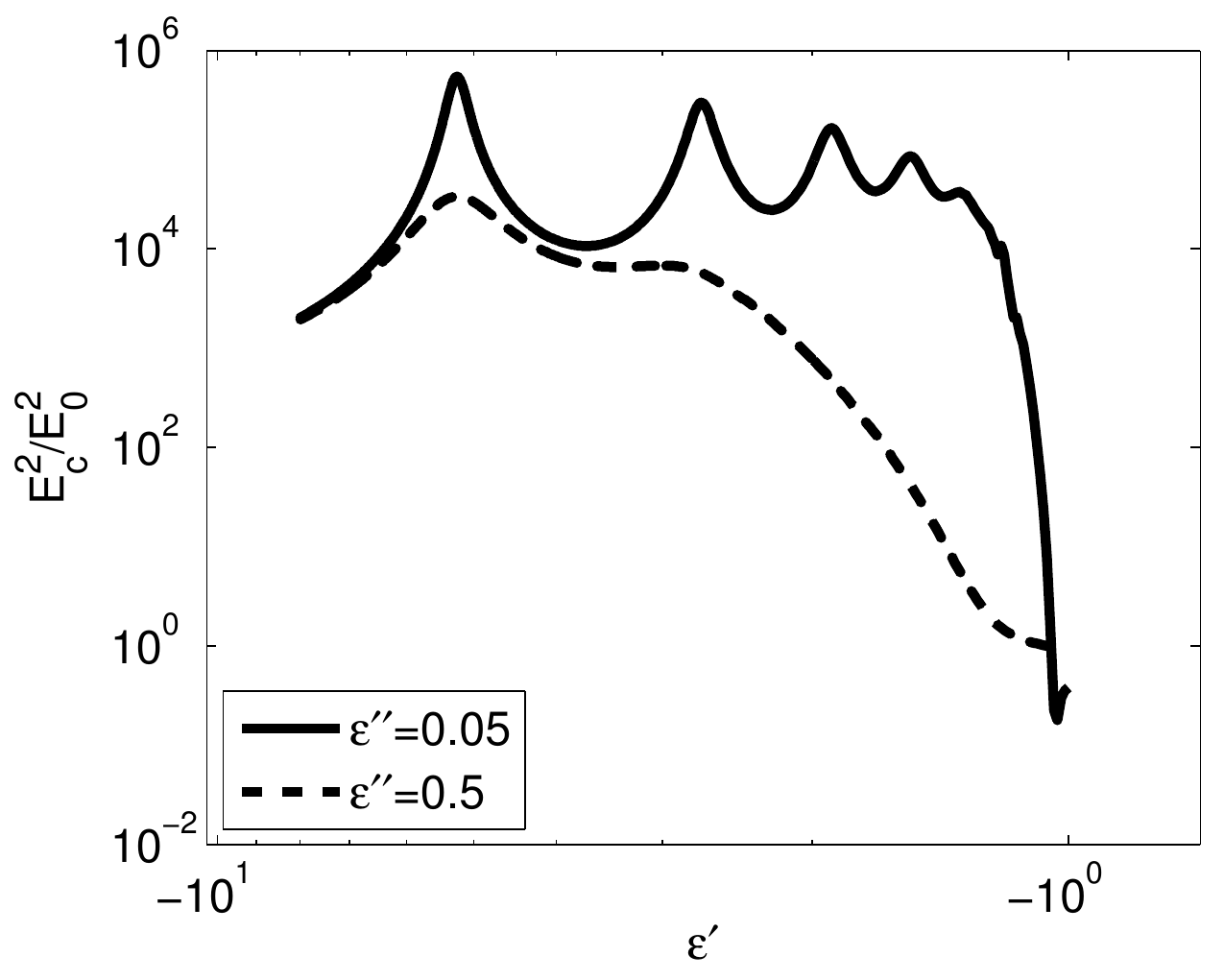}
\caption{
Numerical simulations results for electric field enhancement in the center of the gap between the
cylinders as a function of real part $\varepsilon^{\prime}$ of model metal permittivity.
Wavelength is \hbox{$\lambda=2$ \textmu m}, radius of the cylinders is \hbox{$a=50$ nm}, gap width is \hbox{$\delta=2$ nm}.
}\label{fig:resonance-peaks}
\end{figure}

The complete system of Maxwell's equations can be reduced to wave equation on the only
nonzero magnetic field $Z$-component $H$,
\begin{eqnarray}
\mathop{\mathrm{div}}\frac{1}{\varepsilon({\bm r})}\mathop{\mathrm{grad}} H
+\frac{\omega^2}{c^2}H\label{wave_equation}
= 0,
\end{eqnarray}
where $\omega$ is the frequency of the incident wave.
We assume the magnetic permeability $\mu$ to be unity both in metal and in surrounded space.
The electric field is orthogonal to $OZ$-axis,
thus the problem becomes effectively two-dimensional.
The electric field distribution can be restored as
%\textcolor{red}
{$E^\alpha = (ic/\omega\varepsilon({\bm r}))\eta_{\alpha\beta}\partial_\beta H$,
where $\eta_{\alpha\beta}$ is two-dimensional antisymmetric tensor,
$\eta_{xy}=1$, and $\alpha,\beta$ runs values $x,y$.}

The numerical simulations were performed with commercially available software JCMsuite \cite{Sven}.
The software solves Maxwell's equations based on a finite element method (FEM).
It gives high advantages in simulations of structures with small curved elements.
In particular, the package JCMsuite showed very good results in a benchmark simulation of plasmonic nano antennas \cite{Solvers-1},
which are similar to our structure.
The simulations are challenging because of the narrow gap between cylinders.
The studied system is under resonance condition and exhibits large field enhancement,
so field distribution based adaptive meshing should be applied.
An optimal number of refinement steps was found as well as a number of points on circuits in the vicinity of the gap for the manual mesh specification.
A solution convergence based on posteriori error estimation was examined in the same way as in \cite{Sven} and \cite{Solvers-2}.

For the modes under consideration,
the magnetic field $H$ is symmetric with respect to the axis $OX$ of the system
and the electric field lines are normal to the axis.
The fact allowed us to choose the computational domain
which contains one half of the system shown in FIG. \ref{chain2} and
perfect metal boundary conditions at the $OX$ axis.

Setting permittivity of surrounding medium $\varepsilon_\mathrm{d}=1$ in numerical simulations,
we suppose that permittivity ratio is equal to complex metal permittivity and can be expressed as $\varepsilon=\varepsilon^\prime+i\varepsilon^{\prime\prime}$, i.e.  $\varepsilon^\prime$ is the real part of the permittivity and $\varepsilon^{\prime\prime}$ is the imaginary part.

%\textcolor{red}
{In the following we use the term \textquoteleft model\textquoteright\ metal. By this we mean that we can arbitrarily
assign any values to its permittivity at any wavelength in order to illustrate the dependence of certain
quantities (eg. resonance conditions, field enhancement factor) on the parameters of the system. Having understood
the general properties of plasmon modes in such systems we turn to investigation of the optical properties
on silver cylinders. For silver, both the real $\varepsilon^{\prime}$ and
the imaginary $\varepsilon^{\prime\prime}$ parts of the dielectric permittivity
are functions of the wavelength $\lambda$, which dependencies can be extracted from  \cite{JohnsonKristy}}.

\section{Analytic solution}
\label{sec:analytic_solution}

If the cylinders
radius $a$ is much less than the wavelength in the outer space,
$\sqrt{\varepsilon_\mathrm{d}}ka\ll 1$,
and the skin layer in the metal, $\sqrt{|\varepsilon_\mathrm{m}|}k a\ll 1$,
with $k$ being the free space wavenumber,
it is possible to develop an analytical solution. One can neglect
retardation effects and assume $\omega=0$ in Eq. (\ref{wave_equation}),
thus describing the system in the quasi-static approximation as a pair of cylinders
in a homogenous external electric field. The solution of such problem depends on frequency implicitly via metal permittivity $\varepsilon(\omega)$. This description is valid at distances close to the system, $r\ll \lambda/\sqrt{\varepsilon_\mathrm{d}}$.

\begin{figure}[h]
\includegraphics[width=0.95\textwidth]{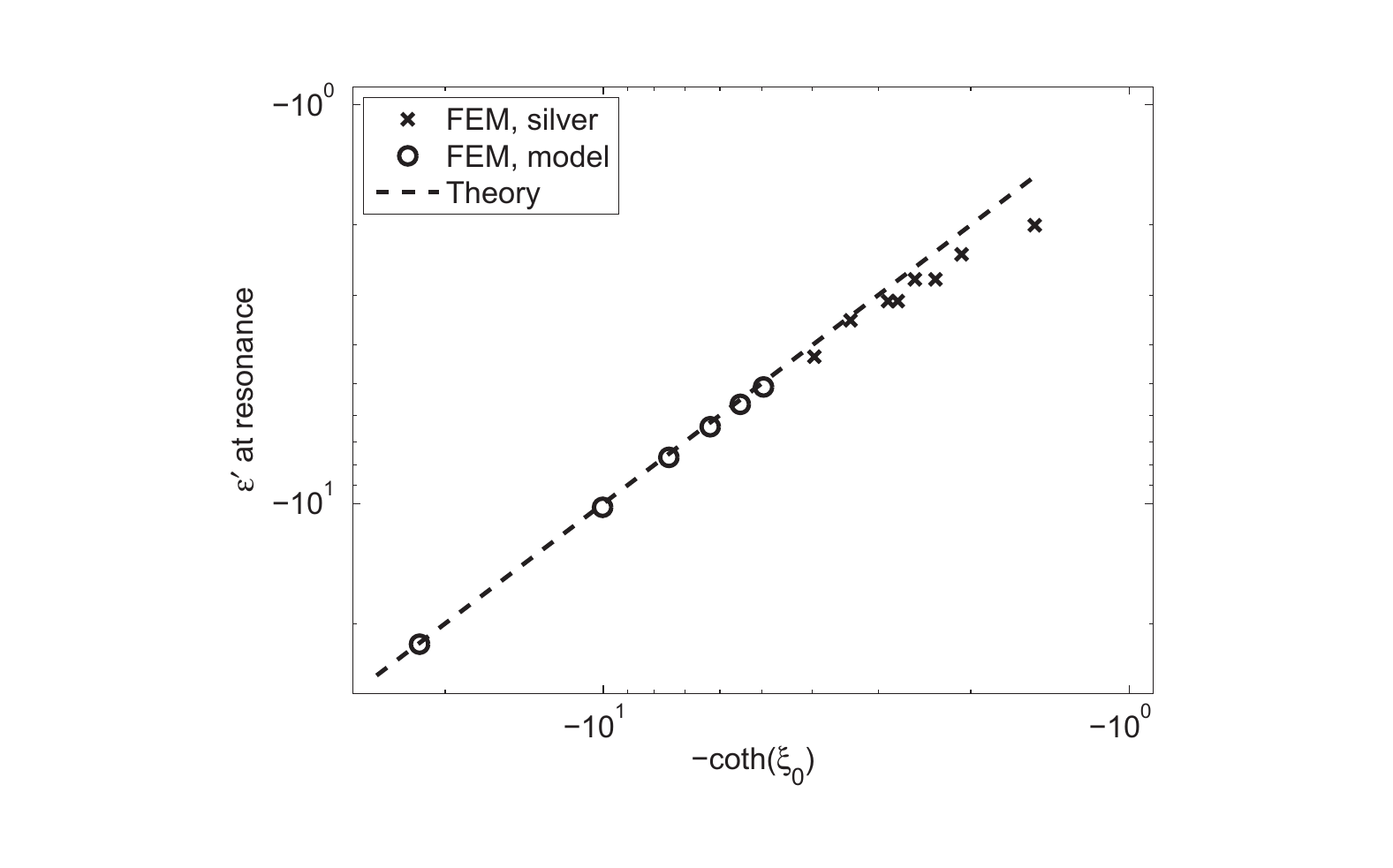}
\caption{
Value of real part of permittivity $\varepsilon$ as
a function of affinity of the cylinders for main resonance, $n=1$.
It was taken \hbox{$a=50$ nm},
$\varepsilon^{\prime\prime}=0.6$ and \hbox{$\lambda=2$ \textmu m}
in numerical simulations for model metal.
For silver $a=15$nm was taken.
}
\label{fig:varepsilon1}
\end{figure}

\begin{figure}[t9]
\includegraphics[width=0.9\textwidth]{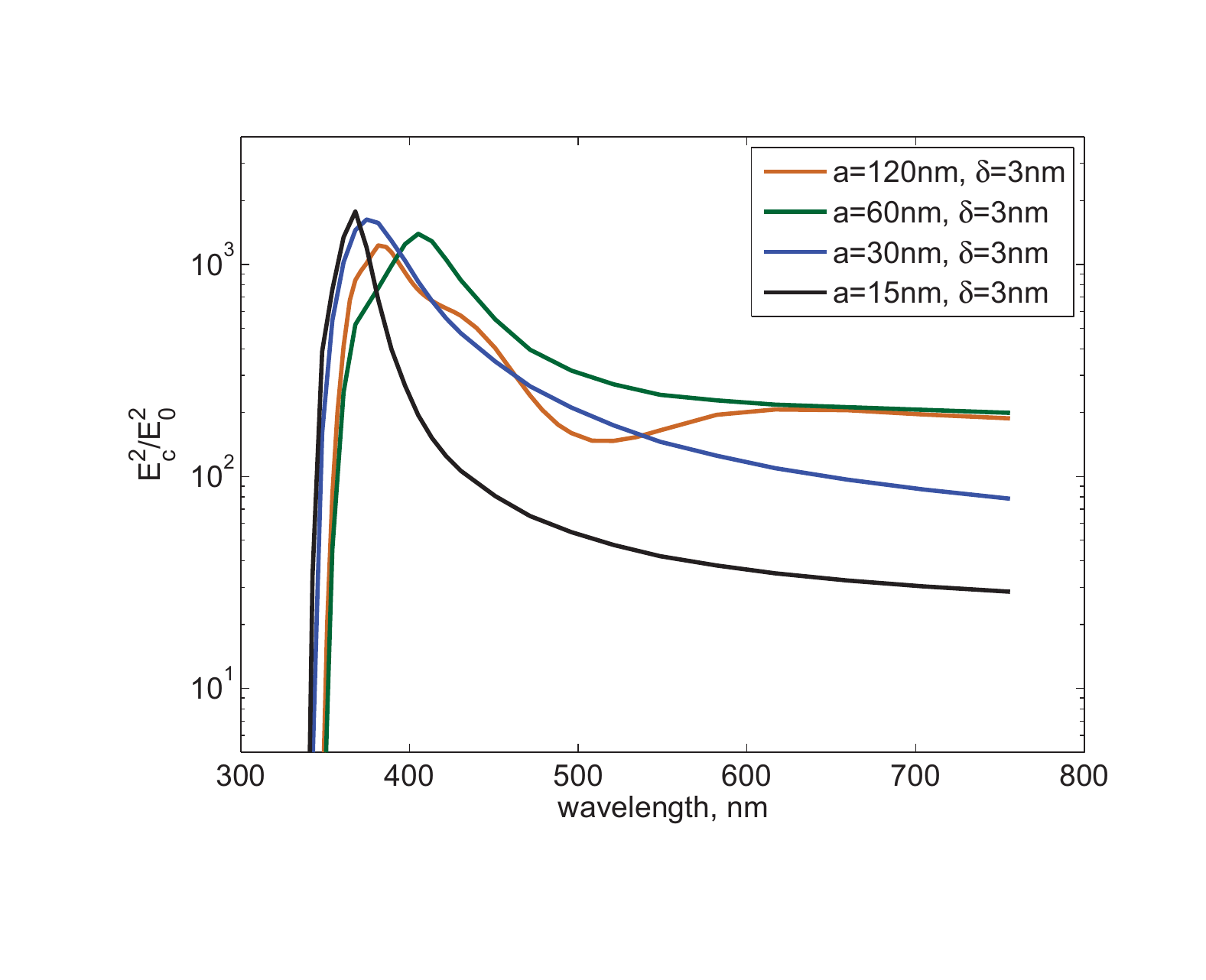}
\caption{
Numerical simulations results for electric field enhancement in the center of the gap for silver cylinders
as a function of wavelength for constant width gap $\delta=3$ nm.
Compare with FIG.~\ref{fig:resonance-peaks}.
}
\label{fig:resonance-peaks_silver_delta}
\end{figure}

\begin{figure}[t9]
\includegraphics[width=0.9\textwidth]{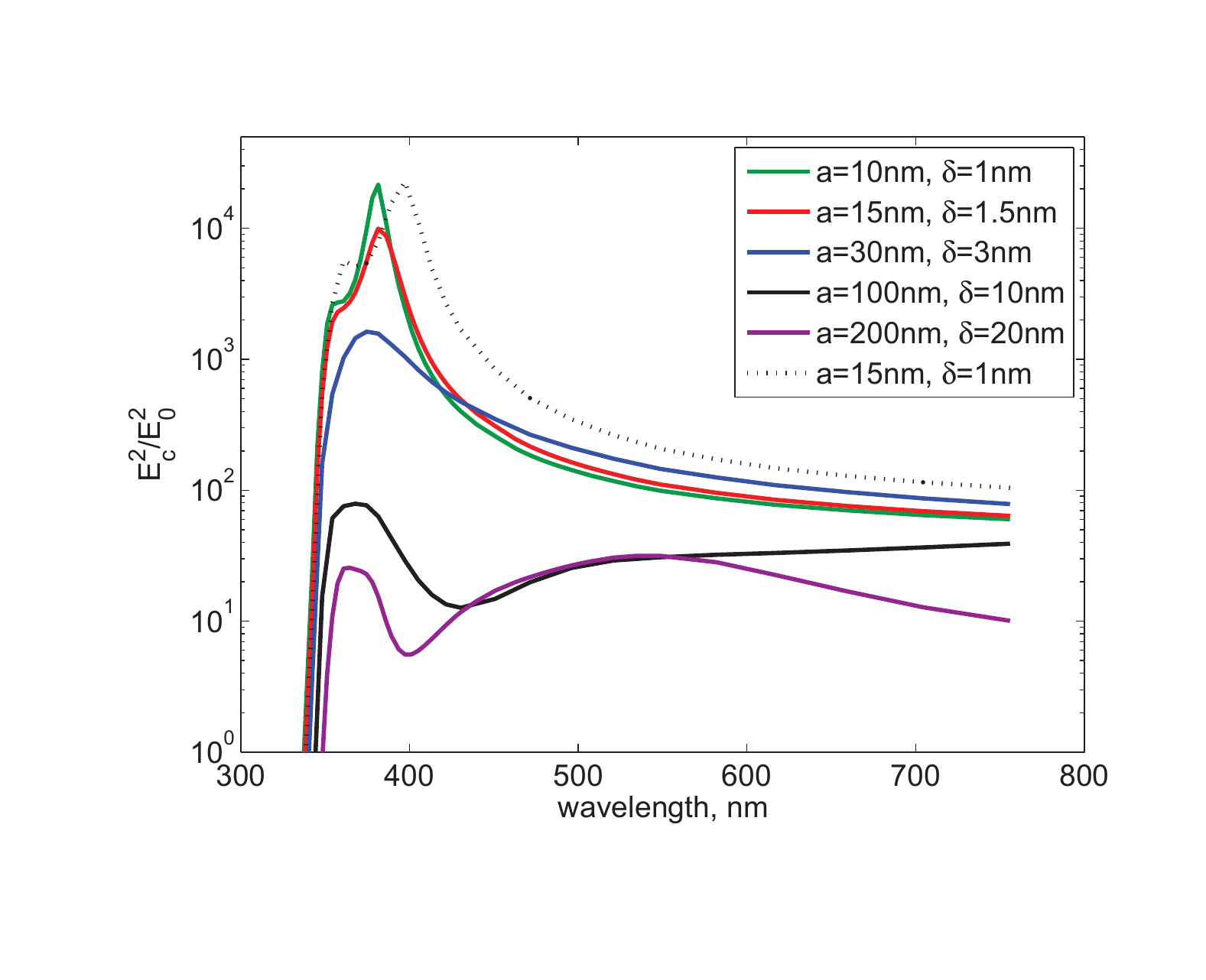}
\caption{
Numerical simulations results for electric field enhancement in the center of the gap for silver cylinders
as a function of wavelength for constant ratio $a/\delta=10$ (solid lines).
Dashed line is given for compasiron and corresponds to different ratio $a/\delta=15$.
Compare with FIG. \ref{fig:resonance-peaks}.
}
\label{fig:resonance-peaks_silver_xi0}
\end{figure}

It is more convenient to use the electric field potential $\Phi$
instead of the magnetic field $H$ in this approximation. The electric field is ${\bf
E}=-\mathop{\mathrm{grad}}\Phi$ in the case. The reason is that the
qualitative physical picture of the surface plasmon oscillations has
clear explanation in terms of the potential $\Phi$. Maxwell's
equations are reduced to quasi-static equation
\begin{eqnarray}\label{equation_potential}
\mathop{\mathrm{div}}\left(\varepsilon({\bm
r})\mathop{\mathrm{grad}} \Phi\right) = 0
\end{eqnarray}
in the vicinity of the cylinders.
Equation (\ref{equation_potential}) reduces to Laplace equation
inside and outside the cylinders: $\Delta \Phi=0$. The boundary conditions
 are the continuity of the potential and normal component
of the electric displacement $\varepsilon({\bm r})\partial_n\Phi$.
The potential should tend to unperturbed external field potential
far from the system, thus $\Phi\to -E_0y$ when $r\gg a$, where $E_0$ is the electric field of the incident wave.
Note here, that the symmetry of the external field potential $-E_0y$ in respect to $OXZ$-plane
corresponds to the symmetry of potential in surface plasmons which are realized at high negative permittivity ratio $\varepsilon$.

\begin{figure}[h]
\includegraphics[width=0.8\textwidth]{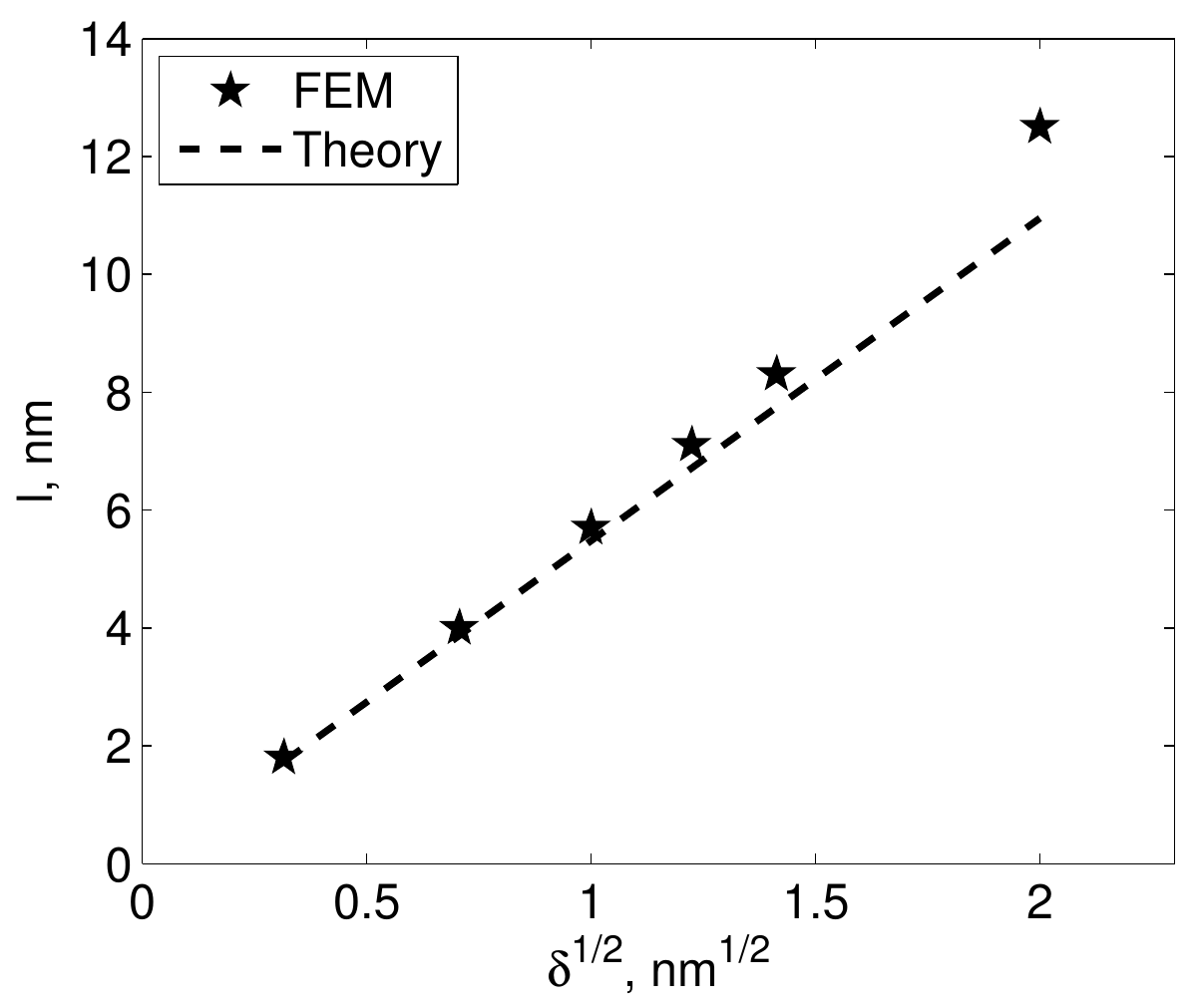}
\caption{
Dependence of mode size on geometrical parameters, for \hbox{$\lambda=2$ \textmu m},
$\varepsilon^{\prime\prime}=0.6$, \hbox{$a=30$ nm}.
Deviation from the dashed line corresponding to law (\ref{mode-size}) at large $\delta$
is due to contribution from non-resonance harmonics in expansion (\ref{out}).
}
\label{fig:mode-size-graph}
\end{figure}

\begin{figure}[t9]
\includegraphics[width=0.9\textwidth]{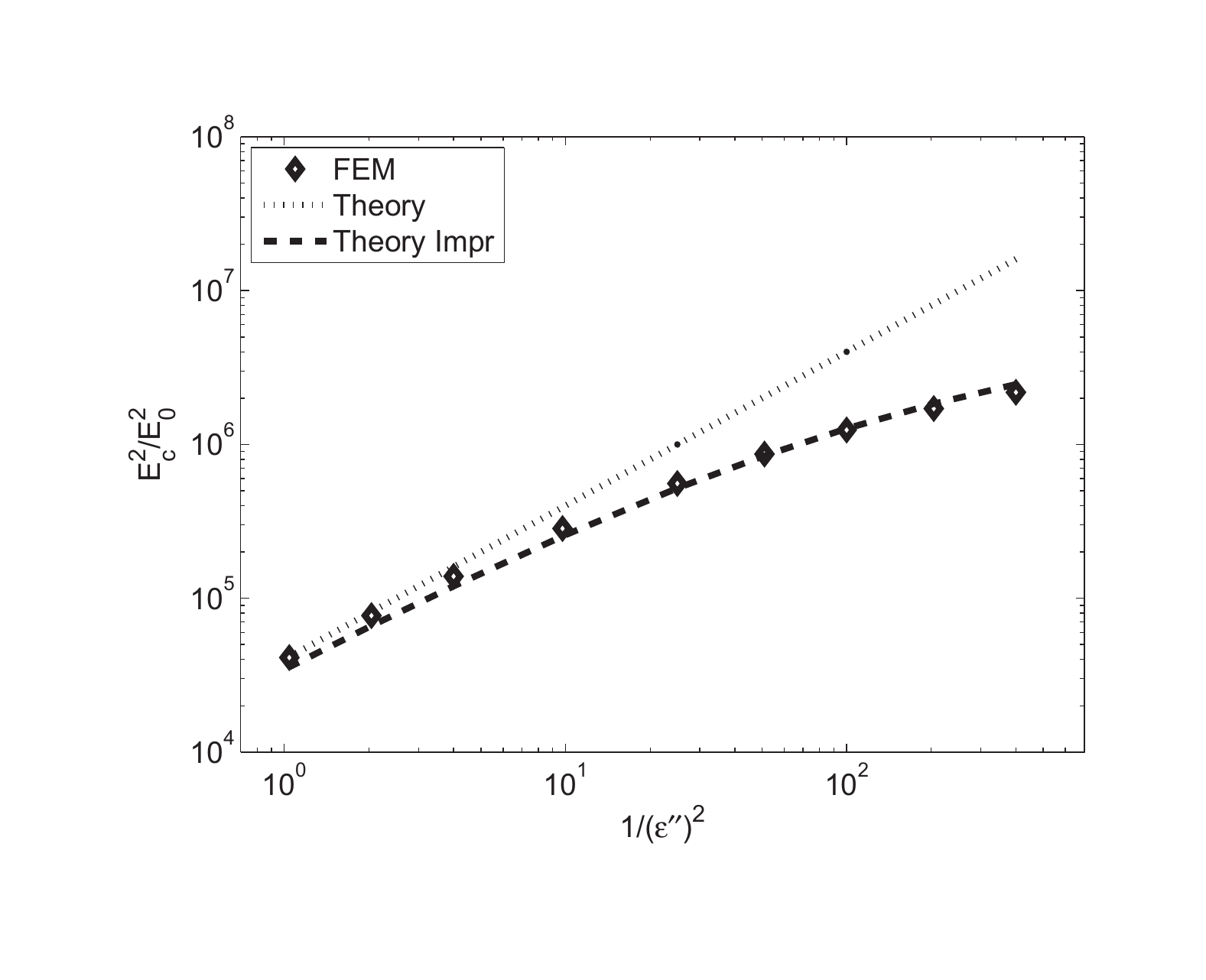}
\caption{
Dependence of electric field enhancement on imaginary part $\varepsilon^{\prime\prime}$ of
model metal permittivity in the first resonance.
Wavelength is \hbox{$\lambda=2$ \textmu m}, radius of the cylinders is \hbox{$a=50$ nm}, gap width is $\delta=1$ nm.
Dotted line corresponds to law (\ref{Ecenter}) at resonance in pure quasi-static limit,
dashed line accounts for radiative losses which are given by (\ref{varepsilonpp_rad-correction}).
}\label{fig:epsilon2-decay}
\end{figure}

Equation (\ref{equation_potential}) does not explicitly contain dependence on the frequency.
The dependence of the scattering properties of the system stems from the
dependence of the metal permittivity on the frequency.
Thus, the most
%\textcolor{red}
{dependencies} presented in this paper are given in terms of the real part $\varepsilon^{\prime}$
of the metal permittivity.
Knowing dispersion of permittivity for a given metal one can
%\textcolor{red}
{rewrite all the dependencies} in
terms of frequency.
Here we note that the plots are not changed dramatically in the case since
the dependence of the metal permittivity is usually
monotonic in the frequency domain under consideration.

In order to solve the above stated problem we use the so-called
bipolar coordinates system: two dimensionless
coordinates $\xi$ and $\eta$ are related to the Cartesian
coordinates as follows:
\begin{equation}
x=\frac{a\sinh\xi_0\sin{\eta}}{\cosh{\xi}-\cos{\eta}};\qquad
y=\frac{a\sinh\xi_0\sinh{\xi}}{\cosh{\xi}-\cos{\eta}}\label{coordinates}
\end{equation}
The lines $\xi=\pm const$ are the pairs of circles situated
symmetrically in respect to $OX$-axis (see FIG. \ref{chain2}).
By definition $\xi_0$ corresponds to the cylinder surfaces, that is $\sinh^2(\xi_0/2)=\delta/(4a)$.
Reference system transformation (\ref{coordinates}) is a conformal map.
The Laplace operator in bipolar coordinates is given by the
following expression:
\begin{equation}
\nabla^2\equiv \frac{1}{h^2(\xi,
\eta)}\left(\frac{\partial^2}{\partial \xi^2} +
\frac{\partial^2}{\partial \eta^2}\right)\label{LB}
\end{equation}
where $h=a\sinh\xi_0/(\cosh{\xi}-\cos{\eta})$ is the scaling function.

The partial solutions of Laplace equation can be written as $e^{\pm
n \xi}\cos n\eta$ or $e^{\pm n \xi}\sin n\eta$ in separated variables.
In order to solve
the problem of two cylinders in the external field one has to expand
the potential in terms of these partial solutions. Using the
symmetry we can write:
\begin{equation}
\Phi^{in}=E_0 a\sum_{0}^{\infty} A_n e^{-n\xi}\cos{n
\eta};\qquad\xi>0\label{in}
\end{equation}
\begin{equation}
\Phi^{out,ind}=E_0 a\sum_{1}^{\infty} B_n \sinh{n\xi}\cos{n
\eta}\label{out},
\end{equation}
where $\Phi^{in}$ is the potential inside the cylinders and
$\Phi^{out,ind}$ is the induced part of the potential outside the cylinders.
Coefficients $A_n$ and $B_n$ are to be found from the boundary
conditions.

\begin{figure}[h]
\includegraphics[width=0.9\textwidth]{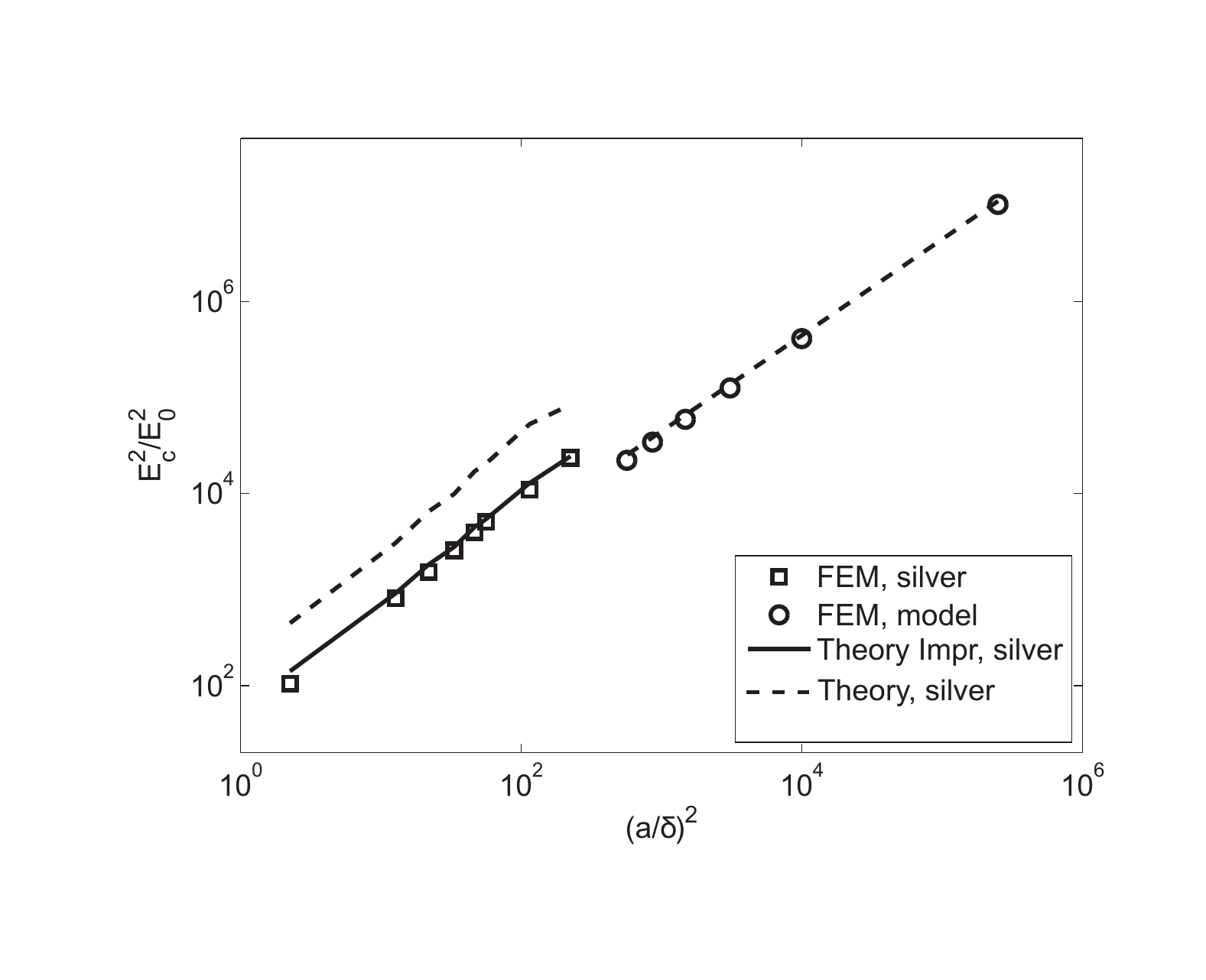}
\caption{
Dependence of field enhancement factor in resonance on geometrical parameters.
It was taken \hbox{$a=30$ nm},
$\varepsilon^{\prime\prime}=0.6$ and \hbox{$\lambda=2$ \textmu m}
in numerical simulations for model metal.
For silver $a=15$nm was taken. Dashed lines correspond to theory prediction~(\ref{Ecenter})
without taking into consideration of radiation losses,
solid line takes the losses into account, see (\ref{varepsilonpp_rad-correction}).
}
\label{fig:Ec2_on_xi0}
\end{figure}

\begin{figure}[h]
\includegraphics[width=0.8\textwidth]{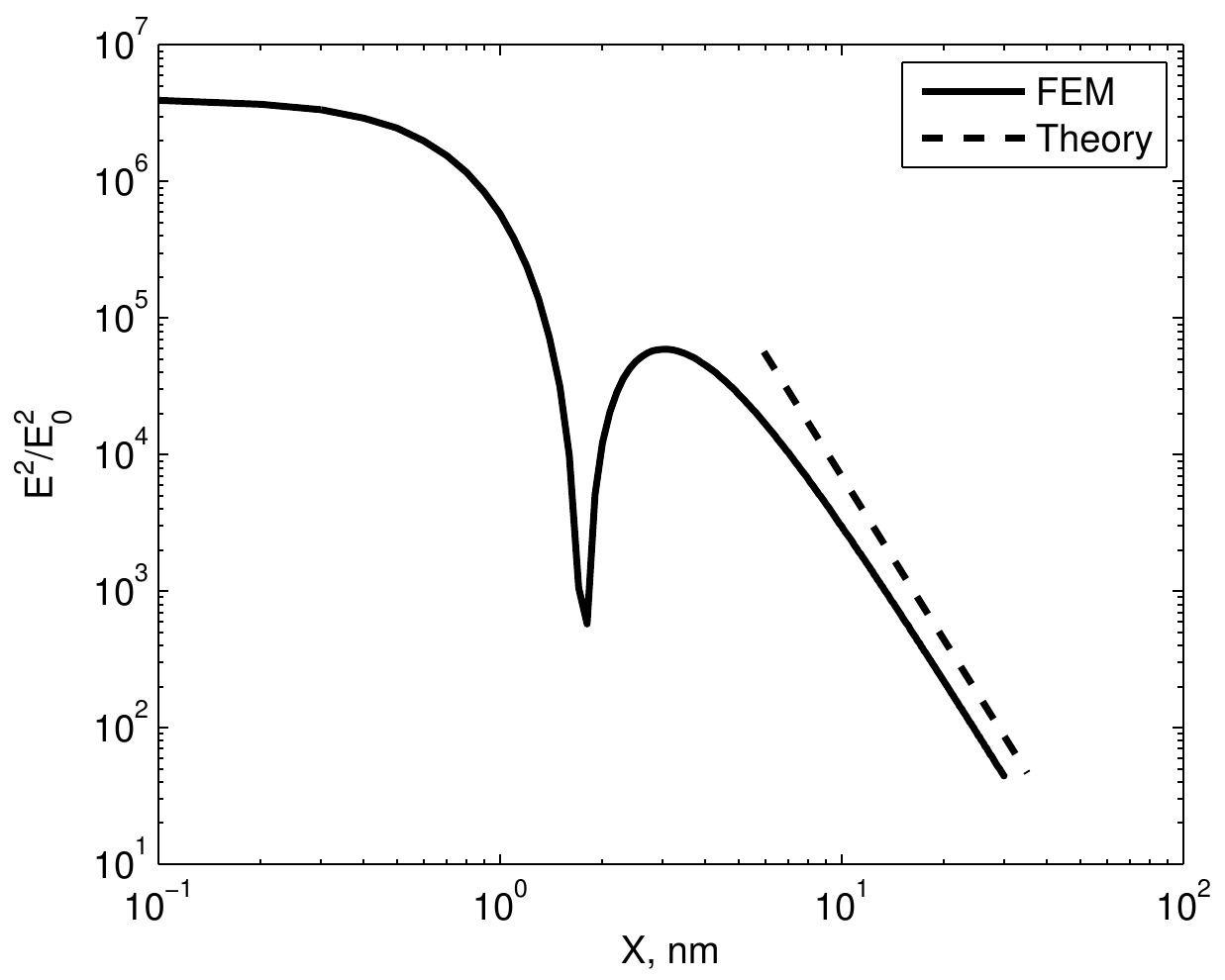}
\caption{
Field spatial dependence on the
middle line between the cylinders. Parameters are $a=30nm$, $\delta=0.1nm$,
$\varepsilon^\prime=-17.4$, $\varepsilon^{\prime\prime}=0.6$, \hbox{$\lambda=2$ \textmu m}.
Dashed line is asymptotics $E\propto1/x^2$, see~(\ref{x2_decay_solar}).
}
\label{fig:x2-decay-graph}
\end{figure}

Now let us expand the potential of the external field
$\Phi_0\equiv -E_0 y$ in terms of partial solutions of Laplace
equation. Simple but cumbersome calculations give:
\begin{equation}
\Phi_0=-\textrm{sign}\xi\; E_0 a
\sinh{\xi_0} \left(\sum_{n=1}^{+\infty} 2 e^{-n|\xi|}\cos{n \eta} +
1\right).
\end{equation}
%\textcolor{red}
{The solution of the problem can be written explicitly ($n\geq 1$)}
\begin{eqnarray}\label{BA_solution}
B_n&=&-\frac{2 \sinh{\xi_0} (1-\varepsilon)e^{-n
\xi_0}}{\sinh{n\xi_0}(\varepsilon+\coth{n\xi_0})}\label{B},
\\[5pt]\nonumber
A_n&=&-\frac{2 \sinh{\xi_0} e^{n
\xi_0}}{\sinh{n\xi_0}(\varepsilon+\coth{n\xi_0})}\label{A}.
\end{eqnarray}
For $n=0$ the
coefficients $B_0=0$, $A_0=-\sinh \xi_0$.

It follows from (\ref{B}) that the resonance occurs when
permittivity contrast $\varepsilon$ takes values
\begin{eqnarray}\label{resonances_position}
\varepsilon_n=-\coth{n\xi_0}\label{resonance}.
\end{eqnarray}

Expression (\ref{out}) allows one to calculate the amplitude of
the electric field.
Explicit expression for electric field components reads
\begin{widetext}
\begin{equation}
E^{ind}_x
=
\bigg(
\sinh\xi\sin\eta
\frac{\partial }{\partial\xi}
-
(1-\cosh\xi\cos\eta)
\frac{\partial }{\partial\eta}
\bigg)
\frac{\Phi^{ind,out}}{a\sinh\xi_0}
\end{equation}
\begin{equation}\label{E_y_general}
E^{ind}_y
=
-
\bigg(
(1-\cosh\xi\cos\eta)
\frac{\partial }{\partial\xi}
+
\sinh\xi\sin\eta
\frac{\partial }{\partial\eta}
\bigg)
\frac{\Phi^{ind,out}}{a\sinh\xi_0}
\end{equation}
\end{widetext}
In what follows, we use these expressions to evaluate the electric field spatial distribution in the vicinity of the cylinders.

\section{Results for the limit of close cylinders}

In this section we analyze the limit of two close cylinders,
when the width $\delta$ of the gap between the cylinders is small compared to
the cylinder radius $a$, i.e. $\delta\ll a$.
The condition means that the dimensionless parameter $\xi_0\approx \sqrt{\delta/a}\ll 1$
in (\ref{coordinates}).

The resonance condition (\ref{resonance}) becomes
\begin{equation}
\varepsilon_{res} = -\frac{1}{n}\sqrt{a/\delta}\label{perm}
\end{equation}
with the limit
$n\lesssim \sqrt{a/\delta}$.
The result can be qualitatively explained as follows.
The resonance condition enables the existence of standing surface plasmon waves in
the flat region of the gap which has the approximate length $\sim \sqrt{a\delta}$
\cite{Kaminov-Mammel_1974_ApplOpt}.
This is a general point,
which is valid in the case of two close
metal spheres as well \cite{Lebedev,Klimov2007}.
Expression (\ref{perm}) is valid until the retardation effects
become important at the scale $\sim\sqrt{a \delta}$,
thus the applicability condition is $\sqrt{\varepsilon_\mathrm{m}} k \sqrt{a\delta} \ll 1$.

Numerical simulations results for positions of the resonances
are shown on FIG. \ref{fig:resonance-peaks} for two different levels of Ohmic losses in metal.
One can see maxima, which correspond to the resonances  (\ref{resonances_position}).
The dependence of the position of the first resonance
on the geometrical parameters of the system
is plotted on FIG.~\ref{fig:varepsilon1},
expression $\coth\xi_0$ tends to $(a/\delta)^{1/2}$ at small gap width.

We performed numerical simulations for silver cylinders
using experimental values of permittivity \cite{JohnsonKristy} for silver.
We chose the size of the system to be close and slightly below
experimentally achieved (see, e.g. \cite{Bakker2008}).
Resonance permittivity (its real part) ratio as the function of geometrical parameters of the system is plotted on
(see FIG.~\ref{fig:varepsilon1})
Electric field enhancement in the center of the gap between
the cylinders as a function of incident wavelength
is plotted on FIG.~\ref{fig:resonance-peaks_silver_delta}, \ref{fig:resonance-peaks_silver_xi0}.
%\textcolor{red}
{On FIG.~\ref{fig:resonance-peaks_silver_delta} all curves correspond to the same value of the gap width $\delta=3$ nm.
On FIG.~\ref{fig:resonance-peaks_silver_xi0}, all curves excepting that for $a=15$ nm, $\delta=1$ nm
correspond to the same value of the ratio $\delta/a=0,1$.
When comparing the plots from these figures with FIG.~\ref{fig:resonance-peaks}, one should bear in mind that the absolute value of the real part $|\varepsilon^\prime|$
of the dielectric permittivity of silver is monotonically increasing function of the wavelength $\lambda$.
One can see, that at most only first two distinct resonance peaks can be observed for silver cylinders, due to the presence of Ohmic losses which lead to broadening and overlapping of all the rest peaks.
Total losses increase with the size of the system, due to increase of radiation intensity.
In FIG.~\ref{fig:resonance-peaks_silver_xi0},
the local minimum near 430 nm and 400 nm of the enhancement factor in the case of cylinders
with radii 100 nm and 200nm correspondingly should be ascribed to retardation effects.
The same concerns the local minimum near 500 nm for the curve corresponding to $a=120$ nm in FIG.~\ref{fig:resonance-peaks_silver_delta}.
For thinner cylinders, retardation  does not lead to any qualitative effects.
In fact, the form of the curve $E_c^2/E_0^2$ as a function of $\lambda$
is independent of the absolute values of the cylinders radii $a$ and the gap thickness $\delta$ in the quasi-static limit.
It is a function of the ratio $a/\delta$ only, that is of the parameter $\xi_0$.
This can be perceived from the general properties of Laplace equation,
which solutions do not change after rescaling of the whole system.
The curves are shifted to lower values of $E_c^2/E_0^2$ for large values of the cylinders' radii
due to radiation losses, the relative importance of which decreases with the wavelength $\lambda$
(see (\ref{varepsilonpp_rad-correction}) below).}

Now let us examine in details the enhancement of the electric field
in the gap when the real part of permittivity contrast $\varepsilon^\prime$ is close to the
first resonance position $\varepsilon_1=-\sqrt{a/\delta}$.
One can see that in the vicinity of the resonance
the contribution from the first harmonics in (\ref{out}) is much
larger than from all the rest.
It follows from the symmetry of the problem that electric field on
the $OXZ$-plane is directed normal to the plane
and is equal to
\begin{eqnarray}\label{x2-decay-formula}
E_y =
\frac{4(a/\delta)E_0}{\varepsilon - \varepsilon_1}
\frac{x^2/a\delta-1}{((x^2/a \delta)^2+1)^2}
\end{eqnarray}
where we have taken into account the fact that $\delta \ll a$.
This expression is valid provided $kx\ll1$.
Electric field changes its sign at $x = l$, where
\begin{equation}\label{mode-size}
l=\sqrt{a\delta},
\end{equation}
and this length should be interpreted as the mode size,
i.e. wavelength of plasmon in
flat dielectric gap with thickness $\delta$ between two bulk metal media
\cite{Kaminov-Mammel_1974_ApplOpt}.

Numerical simulation results for
the mode size as a function of geometrical parameters
of the system are shown in FIG. \ref{fig:mode-size-graph}.
The expression (\ref{mode-size})
is consistent with our qualitative explanation (\ref{perm})
and is valid under the same condition $\sqrt{\varepsilon_\mathrm{m}}
k \sqrt{a\delta} \ll1$.

It follows from (\ref{x2-decay-formula})
that the field at the center of the gap is
%\textcolor{red}
{\begin{equation}
	E_c
	=
	-\frac{4}{\sinh^2\xi_0}\frac{E_c}{\varepsilon-\varepsilon_1}
	=
	-\frac{4a}{\delta}
	\frac{E_0}{\varepsilon - \varepsilon_1}
\label{Ecenter}
\end{equation}
The first representation in (\ref{Ecenter}) is written for general case, whereas the second one is valid for
the limit of closely located cylinders.}
At the resonance, i.e. real part $\varepsilon^\prime=\varepsilon_1$,
Eq. (\ref{Ecenter}) is reduced to $E_c=-4 E_0(a/\delta)/\varepsilon^{\prime\prime}$.
The field enhancement at the resonance as a function of imaginary part of the permittivity
contrast is shown in FIG. \ref{fig:epsilon2-decay}.
The deviation of the enhancement factor in numerical simulations from the analytical prediction
(\ref{Ecenter}) at small $\varepsilon^{\prime\prime}$ is due to radiation losses which become
important for these values of $\varepsilon^{\prime\prime}$, whereas
law (\ref{Ecenter}) accounts for Ohmic losses in metal only.
The radiation losses
are determined mainly by the dipole radiation since the size of the
system is much less than the wavelength.
The dipole moment of the system (per unit length) $d \sim a\delta E_c$,
can be extracted from far asymptotics of the electric field
(\ref{x2-decay-formula}). Radiation losses per unit length
can be estimated as $I\sim\omega^3 d^2/c^2$.  The Ohmic losses in metal can be estimated by multiplying the
dissipation ratio per unit volume $\sim \varepsilon^{\prime\prime}(E_c/\varepsilon^{\prime})^2$ by the volume, occupied by the field $\sim a\delta$.
Thus the dissipation rate (per unit length of cylinders) is estimated as
$Q\sim \varepsilon^{\prime\prime} E_c^2 \delta^2$. Expression (\ref{Ecenter}) is valid provided $I\ll Q$, that is while
$\varepsilon^{\prime\prime}\gg (ka)^2$.

The field enhancement at the resonance
in the gap between the cylinders as a function of geometrical parameters
is  plotted on FIG. \ref{fig:Ec2_on_xi0}.
For model metal we assume wavelength of $\lambda=$\textmu m, which means that
the radiation losses are small compared to the Ohmic ones.
For silver cylinders the deviation of the enhancement factor from
law (\ref{Ecenter}) is substantial. This is due to the fact that wavelength at the resonance frequency is
not well above the size of the system.

%\textcolor{red}
{At scales $x\gg\sqrt{a\delta}$} expression
(\ref{x2-decay-formula}) gives quadratic decay law of the electric field as a function of coordinate
\begin{equation}\label{x2_decay_solar}
E_y = -\frac{a\delta}{x^2}E_c.
\end{equation}
The qualitative explanation of this law for scales
%\textcolor{red}
{$\sqrt{a\delta}\ll x\ll a$} is as follows.
The potential difference
$\Delta\Phi$ between the surfaces of the cylinders is constant at the
scales since surface plasmon wavelength in the gap of thickness
$y$ is much larger than the typical $x$.
Thus $E_y = \Delta\Phi/\Delta y$, where $\Delta y \approx x^2/a$ is the gap
width at given $x$. The potential difference can be related to
the electric field strength $E_c$ at the center of the gap through
the condition that the full surface charge on each cylinder is zero.
The surface charge has different signs inside the flat region at
distances $x<l$ and outside the region (the size of the mode).
The electric field can be
assumed to be uniform in $Y$-direction across the gap, thus
$E_cl\sim-\int_{l}^{a}\mathrm{d}x\Delta\Phi/\Delta y$.
The value of this integral is mostly formed at lower limit and finally one finds
$\Delta\Phi\sim-E_c\delta$.
%\textcolor{red}
{To check the decay law (\ref{x2_decay_solar}) in numerical experiment,
we chose extremely small value of the gap to provide large difference between scales $l$ and $a$,
the result is presented on FIG.~\ref{fig:x2-decay-graph}.}

%\textcolor{red}
{Decaying law (\ref{x2_decay_solar}) for larger scales, $x\gg a$, corresponds to field of dipole
$d = -2a^2 E_0/(\varepsilon - \varepsilon_1)$, the asymptotics is valid for any value of the ratio $\delta/a$.
Note, that the dipole moment is formed at scales $x\sim a$,
and the minor part of the total surface charge is involved in the process.}
In fact, the dipole moment can be estimated as the integral over the region
$l\ll x\ll a$, $d\sim\int \mathrm{d}x \Delta \Phi$.
The integral is formed at distances $x\sim a$,
where only the small part $\sqrt{\delta/a}$ of surface charge is located.
Inner region does not give feasible contribution in $d$ due to the small charge separation length.

%\textcolor{red}
{One can improve Eq. (\ref{Ecenter}) by taking into account the radiation losses.
As a result, the resonance value of dielectric permittivity $\varepsilon_1$
(\ref{resonances_position}) achieves imaginary negative contribution $i\varepsilon_1^{\prime\prime}$, $\varepsilon_1^{\prime\prime}<0$.
The correction can be extracted from the condition, that dipole radiation is equal to power supplied to the system from
external electric field provided Ohmic losses are zero, $I=W$.
The intensity of the radiation is $I=(\pi ck^3/4)|d^2|$, and the power supplied to the system is $W = \omega E_0\mathrm{Im}[d]/2$,
where $E_0$ is assumed to be real and $\mathrm{Im}[d]$ is the imaginary part of the dipole moment induced in the system.
At resonance conditions, dipole moment is pure imaginary, thus we find $d = (2i/\pi)E_0/k^2$.
Comparing the result with (\ref{Ecenter}), we obtain
\begin{eqnarray}\label{varepsilonpp_rad-correction}
	\varepsilon_1^{\prime\prime} = -\pi k^2 a^2,
\end{eqnarray}
where $\varepsilon_1^\prime$ should be determined from (\ref{resonance}).
The correction is valid if $(ka)^2\ll1$ and is applicable for arbitrary ratio of cylinders' radii and the inter-cylinder distance,
including the case when $\delta\sim a$.
In FIG. \ref{fig:epsilon2-decay} and \ref{fig:x2-decay-graph} theoretical curves which take into account the
radiation correction  (\ref{varepsilonpp_rad-correction}) are presented.
The curves fit numerical data quite good, thus the deviation of the data curve from the
theoretical prediction (\ref{Ecenter}) with pure real $\varepsilon_1$
should be assigned just to unaccounted radiation losses.}

\section{Conclusion}

We have numerically investigated the distribution of electro-magnetic field
induced by the incident plane wave near a
system of two closed parallel metallic cylinders of nanoscale dimensions
modeled by Maxwell's equations.
The comparison of the results of the numerical simulation with
the analytical solution for the plasmon modes in the system
governed by Maxwell's equations in the quasistatic limit
showed agreement between these two approaches.

The position of the resonance is determined by the geometrical characteristics of the gap between the cylinders.
Resonance corresponds to existence of standing wave inside the almost flat part of the gap,
which can be thought of as a plane metal-dielectric-metal structure.
Since the wavelength of the plasmon depends on its frequency (through permittivity of metal), there exist
a set of resonance frequencies corresponding to the set of standing modes in the gap.
Comparison with numerical simulations gives a good agreement,
since the dimension of the gap is smaller than the size of the whole metallic granules,
and the quasi-static approximation used in the analytical theory has high accuracy.

%\textcolor{red}
{The analytically obtained \cite{Vorobev}
field enhancement factor in the center of the gap for the first resonance
has also good agreement with that obtained numerically.
It is determined both by Ohmic losses in the metal and radiative losses,
the relative importance of the radiative losses rises with the dimension of the system.}
We assumed the radii of the cylinders to be of the order of 100 nm, which is implemented in experiments \cite{Bakker2008}.
The skin layer depth in metal is of the same order thus the quasi-static approximation has some deviations from the exact
solution of the Maxwell's equations.

Finally, we studied the dependence of the field enhancement for the case of silver cylinders with
permittivity taken from experimental data. Our numerical simulations for such systems showed that
only the Ohmic losses are reasonable, and only the first two resonances can be really observed.

The authors thank V.V. Lebedev and I.R. Gabitov for fruitful discussions and Lin Zschiedrich for assistance with numerical simulations. The work is partially supported by Russian Federal Targeted Programs "S\&S-PPIR",  "I\&DPFS\&T"
and German Academic Exchange Service (DAAD).


\begin{thebibliography}{17}

\bibitem{Suzuki2010}

M. Suzuki, A. Takada, T. Yamada, T. Hayasaka, K. Sasaki, E. Takahashi, S. Kumagai,
\textquotedblleft Low-reflective wire-grid polarizers with absorptive interference overlayers,\textquotedblright\
Nanotechnology {\bf 21}, 175604 (2010)

\bibitem{Ekinci2006}
Y. Ekinci, H.H. Solak, Ch. David, and H. Sigg,
\textquotedblleft Bilayer Al wire-grids as broadband and high-performance polarizers,\textquotedblright\
Optics Express {\bf 14}, 2323 (2006)

\bibitem{Bakker2008}
R.M. Bakker, H-K. Yuan, Z. Liu, V.P. Drachev, A.V. Kildishev, V.M. Shalaev, R.H. Pedersen, S. Gresillon, A. Boltasseva,
\textquotedblleft Enhanced localized fluorescence in plasmonic nanoantennae,\textquotedblright\
Applied Physics Letters {\bf 92}, 043101 (2008)

\bibitem{Lakowicz2007}
J. Zhang, Y. Fu, M.H. Chowdhury, and J. R. Lakowicz,
\textquotedblleft Metal-enhanced single-molecule fluorescence on silver particle monomer and dimer: coupling effect between metal particles,\textquotedblright\
Nano Letters {\bf 7}, 2101 (2007)

\bibitem{Bloemendal2006}
D. Bloemendal, P. Ghenuche, R. Quidant, I. G. Cormack, P. Loza-Alvarez, and G. Badenes,
\textquotedblleft Local Field Spectroscopy of Metal Dimers by TPL Microscopy,\textquotedblright\
Plasmonics {\bf 1}, 41 (2006)

\bibitem{Jain2007}
P.K. Jain, W. Huang, and M.A. El-Sayed,
\textquotedblleft On the Universal Scaling Behavior of the Distance Decay of Plasmon Coupling in Metal Nanoparticle Pairs: A Plasmon Ruler Equation,\textquotedblright\
Nano Letters {\bf 7}, 2080 (2007)


\bibitem{Berthelot2009}
J. Berthelot, A. Bouhelier, C. Huang, J. Margueritat, G. Colas-des-Francs, E. Finot, J-C. Weeber, A. Dereux, S. Kostcheev, H.I.E. Ahrach, A-L. Baudrion, J. Plain, R. Bachelot, P. Royer, G.P. Wiederrecht
\textquotedblleft Tuning of an optical dimer nanoantenna by electrically controlling its load impedance,\textquotedblright\
Nano Letters {\bf 9}, 3914 (2009)


\bibitem{Zhou2010}
Zh.-K. Zhou, M. Li, Zh.-J. Yang, X.-N. Peng, X.-R. Su, Z.-S. Zhang, J.-B. Li, N.-
Ch. Kim, X.-F. Yu, L. Zhou, Zh.-H. Hao, and Q.-Q. Wang
\textquotedblleft Plasmon-Mediated Radiative Energy Transfer across a Silver Nanowire Array via Resonant Transmission and Subwavelength Imaging,\textquotedblright\
ACS Nano {\bf 4}, 5003 (2010)


\bibitem{Maier2003}
S.A. Maier, P.G. Kik, H.A. Atwater, S. Meltzer, E. Harel, B.E. Koel, A.A.G. Requicha
\textquotedblleft Local detection of electromagnetic energy transport below the diffraction limit in metal nanoparticle plasmon waveguides,\textquotedblright\
Nature Materials {\bf 2}, 229 (2003)

\bibitem{Sanders2006}
A.W. Sanders, D.A. Routenberg, B.J. Wiley, Y. Xia, E.R. Dufresne, and M.A. Reed,
\textquotedblleft Observation of plasmon propagation, redirection, and fan-out in silver nanowires,\textquotedblright\
Nano Letters {\bf 6}, 1822 (2006)


\bibitem{Clarkson2011}
J. Clarkson, J. Winans, and P. Facuhet,
\textquotedblleft On the scaling behavior of dipole and quadrupole modes in coupled plasmonic nanoparticle pairs,\textquotedblright\
Optical Materials Express {\bf 1}, 970 (2011)

\bibitem{Romero2006}
I. Romero, J. Aizpurua, G.W. Bryant, and F.J. Garcia De Abajo,
\textquotedblleft Plasmons in nearly touching metallic nanoparticles: singular response in the limit of touching dimers,\textquotedblright\
Optics Express {\bf 14}, 9988 (2006)

\bibitem{Amendola}
V. Amendola, O.M. Bakr, and F. Stellacci,
\textquotedblleft A Study of the Surface Plasmon Resonance of Silver Nanoparticles by the Discrete Dipole Approximation Method:  Effect of Shape, Size, Structure, and Assembly,\textquotedblright\
Plasmonics {\bf 5}, 85 (2010)

\bibitem{Cheng2011}
Y. Cheng, M. Wang, G. Borghs, and H. Chen,
\textquotedblleft Gold nanoparticle dimers for plasmon sensing,\textquotedblright\
Langmuir: the ACS journal of surfaces and colloids {\bf 27}, 7884 (2011)

\bibitem{Haran2010}
G. Haran,
\textquotedblleft Single-molecule Raman spectroscopy: a probe of surface dynamics and plasmonic fields,\textquotedblright\
Accounts of Chemical Research {\bf 43}, 1135 (2010)

\bibitem{Meshbach}
P.M. Morse and H. Feshbach, {\it Methods of theoretical physics. Part II}
(McGraw-Hill, New York, 1953)

\bibitem{Boardman1977}
A.D. Boardman, and B.V. Paranjape,
\textquotedblleft The optical surface modes of metal spheres,\textquotedblright\
Journal of Physics F: Metal {\bf 7}, 1935 (1977)



\bibitem{Fedyanin2010a}
D.Yu. Fedyanin, A.V. Arsenin, V.G. Leiman and A.D. Gladun,
\textquotedblleft Backward waves in planar insulator–metal–insulator waveguide structures,\textquotedblright\
Journal of Optics {\bf 12}, 015002 (2010)

\bibitem{Kaminov-Mammel_1974_ApplOpt}
I.P. Kaminow, W.L. Mammel, and H.P. Weber,
\textquotedblleft Metal-Clad Optical Waveguides: Analytical and Experimental Study,\textquotedblright\
Applied Optics {\bf 13}, 396 (1974)

\bibitem{Pfeiffer1974}
C.A. Pfeiffer, E.N. Economou, and K.L. Ngai,
\textquotedblleft Surface polaritons in a circularly cylindrical interface: surface plasmons,\textquotedblright\
Physical Review B {\bf 10}, 3038 (1974)

\bibitem{Stockman2004}
P. Nordlander, C. Oubre, E. Prodan, K. Li, and M. I. Stockman,
\textquotedblleft Plasmon Hybridization in Nanoparticle Dimers,\textquotedblright\
Nano Letters {\bf 4}, 899 (2004)

\bibitem{Zhukovsky2011}
S.V. Zhukovsky, C. Kremers, and D.N. Chigrin,
\textquotedblleft Plasmonic rod dimers as elementary planar chiral meta-atoms,\textquotedblright\
Optics Letters {\bf 36}, 2278 (2011)


\bibitem{Hentschel2011}
M. Hentschel, D. Dregely, R. Vogelgesang, H. Giessen, and N. Liu,
\textquotedblleft Plasmonic oligomers: the role of individual particles in collective behavior,\textquotedblright\
ACS Nano {\bf 5}, 2042 (2011)


\bibitem{Petschulat2008}
J. Petschulat, C. Menzel, A. Chipouline, C. Rockstuhl, A. Tuennermann, F. Lederer, T. Pertsch,
\textquotedblleft Multipole approach to metamaterials,\textquotedblright\
Physical Review A {\bf 78}, 043811 (2008)


\bibitem{Chigrin2011}
D.N. Chigrin, C. Kremers, and S.V. Zhukovsky,
\textquotedblleft Plasmonic nanoparticle monomers and dimers: From nano-antennas to chiral metamaterials,\textquotedblright\
Applied Physics B {\bf 105}, 81 (2011)


\bibitem{Lebedev}
V. Lebedev, S. Vergeles, and P. Vorobev,
\textquotedblleft Giant enhancement of electric field between two close metallic grains due to plasmonic resonance,\textquotedblright\
Optics Letters {\bf 35}, 640 (2010)

\bibitem{Klimov2007}
V.V. Klimov, and D.V Guzatov,
\textquotedblleft Strongly localized plasmon oscillations in a cluster of two metallic nanospheres and their influence on spontaneous emission of an atom,\textquotedblright\
Physical Review B {\bf 75}, 24303 (2007)

\bibitem{Vorobev}
P.E. Vorobev,
\textquotedblleft Electric field enhancement between two parallel cylinders due to plasmonic resonance,\textquotedblright\
Journal of Experimental and Theoretical Physics {\bf 110}, 193 (2010)

\bibitem{Michaels2000}
A.M. Michaels, J. Jiang, and L. Brus,
\textquotedblleft Ag nanocrystal junctions as the site for surface-enhanced Raman scattering of single rhodamine 6G molecules,\textquotedblright\
J.Phys.Chem. B {\bf 104}, 11965 (2000)

\bibitem{Sven}
J. Pomplun, S. Burger, L. Zschiedrich, F. Schmidt,
\textquotedblleft Adaptive finite element method for simulation of optical nano structures,\textquotedblright\
Physica Status Solidi (b) {\bf 244}, 3419 (2007)

\bibitem{Solvers-1}
J. Hoffmann, C. Hafner, P. Leidenberger, J. Hesselbarth, S. Burger,
\textquotedblleft Comparison of electromagnetic field solvers for the 3D analysis of plasmonic nanoantennas,\textquotedblright\
Proc. SPIE {\bf 7390}, 73900J (2009)

\bibitem{Solvers-2}
S. Burger, R. K\"{o}hle,  L. Zschiedrich, W. Gao,
F. Schmidt, R. M\"{a}rz, Ch. N\"{o}lscher,
\textquotedblleft Benchmark of FEM, waveguide and FDTD algorithms for rigorous mask simulation,\textquotedblright\
Proc. SPIE {\bf 5992}, 599216 (2005)


\bibitem{Davis1964}
M.H. Davis,
\textquotedblleft Two Charged Spherical Conductors in a Uniform Electric Field: Forces and Field Strength,\textquotedblright\
The Quarterly Journal of Mechanics and Applied Mathematics {\bf 17}, 499 (1964)

\bibitem{Love1924}
A.E.H. Love,
\textquotedblleft Some Electrostatic Distributions in two Dimensions,\textquotedblright\
Proceedings of the London Mathematical Society {\bf s2-22}, 337 (1924)

\bibitem{JohnsonKristy}
P.B. Johnson, and R.W. Christy,
\textquotedblleft Optical Constants of the Noble Metals,\textquotedblright\
Physical Review B {\bf 6}, 4370 (1972)







\end{thebibliography}
\end{document}